\documentclass[%
 reprint,
superscriptaddress,
 amsmath,amssymb,
 aps,
prl,
]{revtex4-2}

\usepackage{float}
\usepackage{physics}
\usepackage{graphicx}
\usepackage{dcolumn}
\usepackage{bm}

\usepackage{xr}
\makeatletter

\newcommand*{\addFileDependency}[1]{
\typeout{(#1)}
\@addtofilelist{#1}
\IfFileExists{#1}{}{\typeout{No file #1.}}
}\makeatother

\newcommand*{\myexternaldocument}[1]{%
\externaldocument{#1}%
\addFileDependency{#1.tex}%
\addFileDependency{#1.aux}%
}
\myexternaldocument{main_supplementary_material}

\newcommand{\expV}[2]{\langle #1 \rangle_{#2}}
\newcommand{\varV}[2]{\Delta #1^2_{#2}}
\newcommand{\stdV}[2]{\Delta #1_{#2}}
\newcommand{\ceil}[1]{\left \lceil #1 \right \rceil }

\begin{document}

\title{Speeding Up Quantum Measurement Using Space-Time Trade-Off}

\author{Christopher Corlett}
\affiliation{Quantum Engineering Centre for Doctoral Training, University of Bristol, Tyndall Avenue, Bristol, BS8 1FD, UK}
\affiliation{H.H. Wills Physics Laboratory, University of Bristol, Tyndall Avenue, Bristol, BS8 1TL, UK}

\author{Ieva \v Cepait\.e}
\affiliation{Department of Physics and SUPA, University of Strathclyde, Glasgow, G4 0NQ, UK}

\author{Andrew J. Daley}
\affiliation{Department of Physics and SUPA, University of Strathclyde, Glasgow, G4 0NQ, UK}
\affiliation{Clarendon Laboratory, University of Oxford, Parks Road, Oxford OX1 3PU, UK}

\author{Cica Gustiani}
\affiliation{LIP6, CNRS, Sorbonne Université, 4 Place Jussieu, 75005 Paris, France}

\author{Gerard Pelegr\'i}
\affiliation{Department of Physics and SUPA, University of Strathclyde, Glasgow, G4 0NQ, UK}

\author{Jonathan D.~Pritchard}
\affiliation{Department of Physics and SUPA, University of Strathclyde, Glasgow, G4 0NQ, UK}

\author{Noah Linden}
\affiliation{School of Mathematics, University of Bristol, Fry Building, Woodland Road, Bristol, BS8 1UG, UK}

\author{Paul Skrzypczyk}
\affiliation{H.H. Wills Physics Laboratory, University of Bristol, Tyndall Avenue, Bristol, BS8 1TL, UK}
\affiliation{CIFAR Azrieli Global Scholars Program, CIFAR, Toronto, Canada}

\date{\today}

\begin{abstract}
We present a scheme for speeding up quantum measurement. The scheme builds on previous protocols that entangle the system to be measured with ancillary systems. In the idealised situation of perfect entangling operations and no decoherence, it gives an exact space-time trade-off meaning the readout speed increases linearly with the number of ancilla. We verify this scheme is robust against experimental imperfections through numerical modelling of gate noise and readout errors, and under certain circumstances our scheme can even lead to better than linear improvement in the speed of measurement with the number of systems measured. This hardware-agnostic approach is broadly applicable to a range of quantum technology platforms and offers a route to accelerate mid-circuit measurement as required for effective quantum error correction.
\end{abstract}

\maketitle

\subsection{Introduction}
High-fidelity readout of individual quantum systems is an essential requirement for emerging quantum technologies including precision metrology and quantum computing. Fast measurements are highly desirable, particularly in the context of mid-circuit measurements for quantum error correction where readout duration imposes a significant overhead \cite{Bluvstein_2023, Bruzewicz_2019, PhysRevLett.127.050501, Krinner_2022}, however in practice there is a trade-off between measurement duration and quality. This places a lower bound on how fast a measurement can be performed to achieve a given fidelity. Even in cases where time is not a finite resource, such as the final state readout at the end of a computation, experimental imperfections and qubit dephasing in real systems impose an upper limit on readout time and hence measurement quality. In both scenarios, we seek a protocol able to simultaneously reduce readout duration and enhance the measurement fidelity.

There has been much previous work on how to improve the quality of measurements without simply increasing their duration. They can be split into two categories, passive methods, which correct errors on expectation values, \cite{Geller_2020, Smith_2021, Bravyi_2021, Maciejewski_2020} and active methods, which correct errors on single-shot measurements \cite{dall_2010, PhysRevA.105.012419, Günther_2022, linden2022use, PhysRevA.103.022407, Funcke_2022, PRXQuantum.4.040305}. For applications such as mid-circuit measurements, passive methods are insufficient as we need high quality results on a single-shot. While active methods have been shown to improve the quality of single-shot measurements, there has been little attention paid to how we can speed them up. The work towards faster measurements is a hard engineering challenge and thus far, progress has only been made for specific architectures \cite{PhysRevA.72.042302, Takeda_2024, swiadek2023, Walter_2017, PhysRevLett.99.120502, Shea_2020, PhysRevApplied.10.034040, Dassonneville_2020, PhysRevLett.127.050501, linden2022use}. No thorough analysis has been performed at the abstract, hardware-agnostic level.

In this paper, we propose a scheme for speeding up quantum measurements. Our scheme builds on protocols in previous work \cite{dall_2010, PhysRevA.105.012419, Günther_2022, linden2022use} where a repetition code is used to spread the information by entangling the system to be measured with  ancillary qubits. The key novelty in our work is to harness these protocols to provide a scheme for reducing measurement time while maintaining (or even improving) measurement quality. While the simplest embodiment of our scheme describes using multiple measurement devices, in fact our scheme can also be implemented using only a single measurement device coupled to the target and ancillary qubits. A further impact of reducing measurement time is to suppress decoherence processes, which allows us to outperform the maximum quality imposed by the upper bound on measurement time. The effect of the scheme is to trade-off space (number of ancillary systems)  against the time taken to do the measurement.

A vital aspect of the scheme is that the trade-off is robust against imperfections in its implementation.  We illustrate this by detailed numerical simulations of a natural, simple but characteristic model of noise and decoherence. Crucially, we note that in certain circumstances with realistic noise parameters our scheme can lead to a better than linear improvement in the speed of measurement with the number of systems measured, achieving optimal readout parameters with a modest number of additional qubits.

In summary, our scheme is hardware agnostic and provides a robust, powerful and versatile method for speeding up quantum measurements using ancillary systems.  

\subsection{Case study: Poisson measurement statistics}
In order to present the main idea, we will first focus on state dependent fluorescence qubit measurement which is the primary method used in trapped-ion \cite{Bruzewicz_2019, PhysRevLett.100.200502} and neutral-atom \cite{Bluvstein_2023, graham2023midcircuit, Wintersperger2023, PhysRevLett.119.180503} quantum computing.

Qubit information is encoded in either a ``dark'' or ``bright''  state of an ion or atom. When illuminated by a detection laser, the atom will be projected into one of these states and will then scatter photons. A measurement device (detector) will count the number $k$ of photons detected. We have in mind to count photons for a time $t$. The bright state will scatter many photons with a mean of $\nu_1(t)$ while the dark state will scatter relatively fewer photons with a mean of $\nu_0(t)$ where $\nu_1(t)\geq\nu_0(t)$. To discriminate the bright and dark states we define a threshold value $\eta (t)$ where $\nu_0(t)<\eta (t)<\nu_1(t)$. If the number of photons counted is $k<\eta (t)$ then we assign the outcome $0$ and if $k\geq\eta (t)$ we assign the outcome $1$. From now on we refer to the bright state as $\ket{1}$ and the dark state as $\ket{0}$. In all that follows, we denote $j\in \{0,1\}$.

Let us denote by $\mathcal{P}_{\ket j,t}^{(n)}(k)$ the probability distribution of the random variable $K_{\ket j,t}^{(n)}$, defining the number of photons counted, when we use $n$ detectors for time $t$ and the input state is $\ket j$. Initially we consider $n=1$.

In an idealised situation, the number of photons counted will follow a \emph{Poisson distribution} ${\mathcal{P}_{\ket j,t}^{(1)}(k)= L_{\nu_j(t)}(k)}$, where $L_\omega(k)=\frac{\omega^k}{k!}e^{-\omega}$.  Recall the mean and variance of a Poisson distribution coincide and are both equal, $\expV{K}{L_\omega}=\varV{K}{L_\omega}=\omega$. The two distributions,  $\mathcal{P}_{\ket 0,t}^{(1)}$ and $\mathcal{P}_{\ket 1,t}^{(1)}$, will have an overlap which is the source of error in measurement quality when assigning outcomes; no threshold value can be defined which will give you a perfect measurement by correctly discriminating the two states.

If $k_1$ photons are emitted during time $t_1$ and $k_2$ photons are emitted during a subsequent time $t_2$ such that the total number of photons emitted in time $t=t_1+t_2$ is $k=k_1+k_2$, then by the fundamental convolution property of the Poisson distribution,
${L_{\nu_j(t_1+t_2)}(k)=(L_{\nu_j(t_1)}*L_{\nu_j(t_2)})(k)=L_{\nu_j(t_1)+\nu_j(t_2)}(k)}$,
i.e.~the Poisson distribution convolves to itself. This implies that $\nu_j(t)= \mu_j t$, for a rate $\mu_j$ independent of $t$.  In particular the convolution property means that $\nu_j(2t)= 2\nu_j(t)$.

Although one is generally interested in how good a measurement is for an arbitrary state, a commonly used figure of merit is the distinguishability, as quantified by the \emph{signal-to-noise-ratio} (SNR), of the ${\mathcal{P}_{\ket 0,t}^{(n)}(k)}$ and ${\mathcal{P}_{\ket 1,t}^{(n)}(k)}$ probability distributions. The SNR is
\begin{equation}\label{eqn:snr}
    S\left(\mathcal{P}_{\ket 0,t}^{(n)}, \mathcal{P}_{\ket 1,t}^{(n)}\right) = \frac{2\left|\expV{X}{\mathcal{P}_{\ket 0,t}^{(n)}}-\expV{X}{\mathcal{P}_{\ket 1,t}^{(n)}}\right|}{\stdV{X}{\mathcal{P}_{\ket 0,t}^{(n)}} + \stdV{X}{\mathcal{P}_{\ket 1,t}^{(n)}}}.
\end{equation}
The larger the SNR, the better the distinguishability and the better the measurement quality.

In our case, for the two Poisson distributions, we find
\begin{equation}
    S\left(\mathcal{P}_{\ket 0,t}^{(1)}, \mathcal{P}_{\ket 1,t}^{(1)}\right) = \frac{2\sqrt{t}|\mu_0-\mu_1|}{\sqrt{\mu_0} + \sqrt{\mu_1}}. \nonumber
\end{equation}
Thus the quality of the measurement is proportional to $\sqrt{t}$ and therefore increases with measurement duration. This provides a quantification of the inverse trade-off between the speed and quality of a measurement.

\subsection{Exact space-time trade-off in the idealised Poisson case}
\begin{figure}
    \centering
    \includegraphics[width=\linewidth]{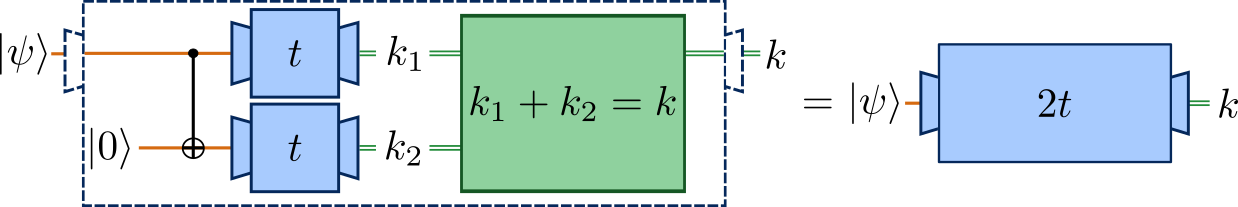}
    \caption{\textbf{Space-time trade-off scheme.}
    Idealised version using one additional ancillary qubit. We entangle the input qubit with an ancillary qubit in state $\ket{0}$ using a CNOT gate. We then measure each qubit for a time $t$, obtaining results $k_1$ and $k_2$ respectively. The statistics of the sum $k = k_1 + k_2$ coincides with the statistics for a single measurement of time $2t$. We thus speed up the measurement by a factor of 2.}
    \label{fig:protocol}
\end{figure}

We now introduce a scheme which, rather than trading time for quality as above, allows us to trade time for ``space'' -- by which we mean number of qubits. In particular, this provides a means to realise high-quality fast measurements, at the cost of using additional ancillas. See Fig.~\ref{fig:protocol}. We apply a CNOT gate between our target qubit and an ancillary qubit initialised in the state $\ket{0}$. In this section we consider the idealised situation that the CNOTs can be performed perfectly and in a negligibly-short time compared to the measurement duration. We measure the ancilla and the original qubit then assign the result as either $0$ or $1$, representing the outcome of the measurement on the target qubit.  In the ideal case, when measured for a time $t$, we will now show that this scheme gives identical measurement statistics to that of a standard measurement on a single qubit measured for a time $2t$. The measurement time is effectively halved.

For now, for clarity of presentation, we analyse the case of independent detectors for each qubit, but come back to this point below. We denote by $k_1$ and $k_2$ the number of photons detected from the first and second qubit detectors respectively and we input the state $\ket{\psi}=\ket{j}$ into our scheme. The measurement statistics for the pair of outcomes $(k_1, k_2)$, given we are measuring the post-CNOT state $\ket{jj}$ for a time $t$, are ${\mathcal{R}_{\ket{jj},t}(k_1, k_2) = L_{\mu_jt}(k_1)L_{\mu_jt}(k_2)}$. The outcomes $(k_1, k_2)$ form a pair of independent random variables, hence the probability distribution of their sum $k=k_1+k_2$ is the convolution of the individual probability distributions, thus ${\mathcal{P}_{\ket j,t}^{(2)}(k) = (L_{\mu_jt}*L_{\mu_jt})(k) = L_{2\mu_jt}(k)}$.
An important feature is present here; since the probability distribution only considers the total sum of emissions from both fluorescing qubits, we can consider the distribution as describing the number of photons counted from a \emph{single detector} which is collecting \emph{all} the emitted photons, rather than collecting separately and then adding.  
 
The key insight is that, due to the convolution property of the Poisson, measuring a single qubit for time $2t$ is equivalent to summing the results from two measurements, of time $t$, on two entangled qubits. Specifically
${\mathcal{P}_{\ket j,t}^{(2)}(k) =  \mathcal{P}_{\ket j,2t}^{(1)}(k).}$
This result shows that we can speed up the measurement, by a factor of $2$, by performing a statistically identical measurement in time $\frac{t}{2}$ as was previously performed in time $t$. Since the measurements are identical, the quality, as quantified by the SNR (or any other figure of merit), will be constant while the measurement is performed faster.

This result generalises to $N$ qubits where we entangle $N-1$ ancillary qubits, in the initial state $\ket{0}$, to the target qubit using CNOT gates. In this case we find 
\begin{equation}\label{eqn:conv}
    \mathcal{P}_{\ket j,t}^{(N)}(k) = (L_{\mu_jt}^{*N})(k) = L_{N\mu_jt}(k) = \mathcal{P}_{\ket j,Nt}^{(1)}(k),
\end{equation}
where $k$ is the total number of photons counted and $(f^{*N})(w)$ is the $N$-fold convolution of the distribution $f(v)$. This shows that we can keep the quality of a measurement constant, and speed it up arbitrarily, from time $t$ to time $t/N$, by making use of $N$ measurements in parallel. 

\begin{figure*}
  \centering
  \includegraphics[width=\textwidth]{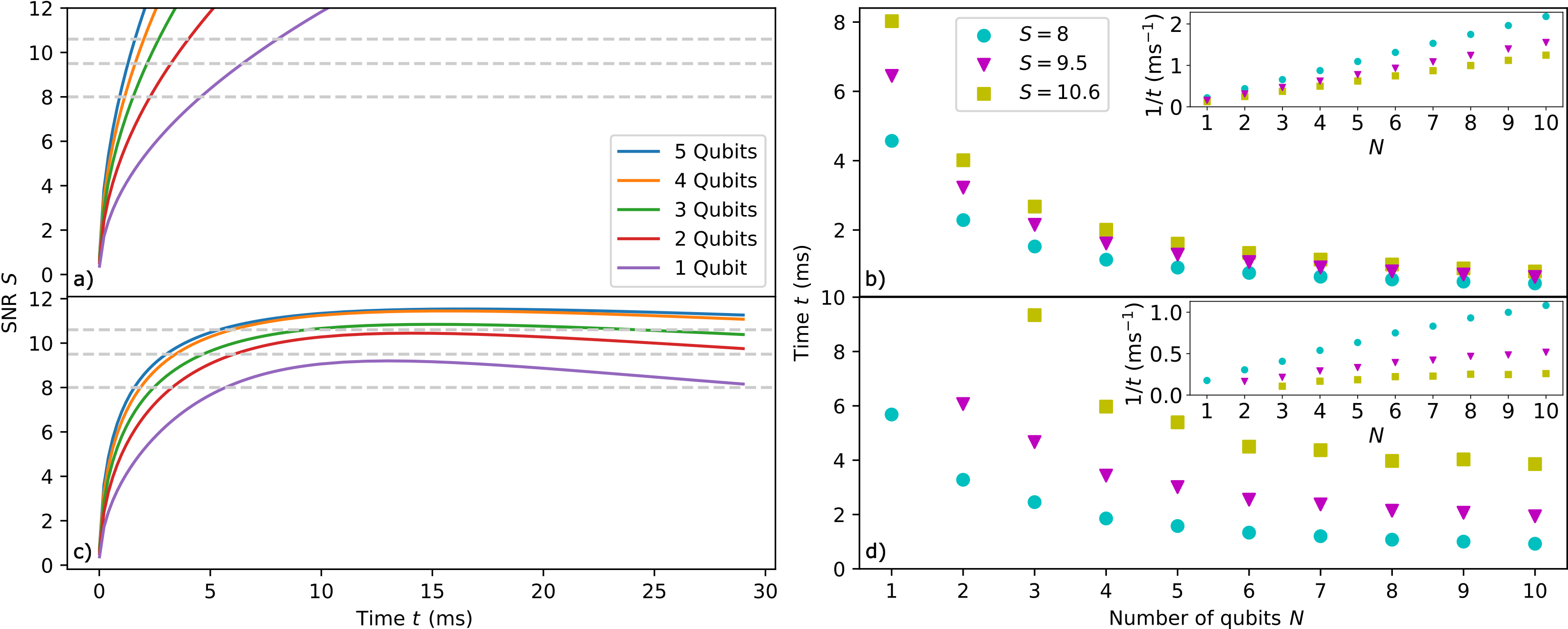}
    \caption{{\bf Space-time trade-off for idealised and realistic devices.} The parameters used for the plots are experimentally relevant and are taken from \cite{PhysRevLett.131.030602}. In plots \textbf{a)} and \textbf{b)}, we use Poisson distributed measurement statistics \eqref{eqn:conv} with $\mu_0=3.5 \text{ms}^{-1}$ and $\mu_1=14\text{ms}^{-1}$ for different numbers of qubits $N$. \textbf{a)} SNR \eqref{eqn:snr} against time from $N=1$ to $N=5$ qubits. The SNR always increases with time and qubit number and continues the trend beyond what is shown on the graph. \textbf{b)} time of measurement $t$ against number of qubits from $N=1$ to $N=10$ qubits for the SNR values on the grey dashed lines in \textbf{a)}. Inset: same points but for the inverse time $1/t$, demonstrating the linear relationship \eqref{eqn:perfect_ratio}. \textbf{c)} and \textbf{d)} show similar results but for realistic devices. In this case, a higher maximum distinguishability can be achieved by introducing more qubits and faster measurements can be made while also improving distinguishability. These plots are generated using the noise and decoherence model \eqref{eqn:noise_and_decay}, found in the Appendix. We use input parameters $\lambda=0.0041\text{ms}^{-1}$ and $p=0.01$ where $p$ is chosen as a conservative estimate of CNOT failure rate. In \textbf{c)}, the SNR increases with time up to a maximum value then decreases due to decoherence. The rate at which the SNR improves increases with qubit number, as does the peak SNR achieved. Inset: points in the inset no longer follow a straight line, indicating that the space-time trade-off is not linear with qubit number $N$.}
    \label{fig:snr_perfect_and_noise}
\end{figure*}

A useful ratio for comparison with more realistic devices later on is $t^{(1)}_S/t^{(N)}_S$ where $t^{(N)}_S$ is the time it takes an $N$ qubit scheme to reach an SNR of $S$. In this case
\begin{equation}\label{eqn:perfect_ratio}
    t^{(1)}_S/t^{(N)}_S=N
\end{equation}
for all $S$, so the improvement in speed is exactly linear in qubit number. The speed up from our scheme is shown in Figs.~\ref{fig:snr_perfect_and_noise}a and \ref{fig:snr_perfect_and_noise}b.

\subsection{General space-time trade-off}
A key insight is that the scheme above applies much more generally than just to Poisson statistics arising from counting photons. In fact, the  scheme  gives a space-time trade-off in a wide range of measuring devices.

We will assume that we have a measurement procedure that is \emph{time dependent}, meaning that the duration of the measurement can be changed, which will change the quality of the measurement. Let us assume that the statistics of this measurement (with outcome $x$) when measuring the two basis states for a time $t$ is given by $\mathcal{P}_{\ket{j},t}^{(1)}(x)$, where the superscript reminds us that this is without using the measurement scheme, i.e.~we are performing a single measurement. For this measurement we can calculate the SNR \eqref{eqn:snr} between these two distributions. Let us assume the dependence with time, to leading order, is 
${S(\mathcal{P}_{\ket{0},t}^{(1)},\mathcal{P}_{\ket{1},t}^{(1)}) \sim t^\alpha.}$
For the case of Poisson statistics above we saw that $\alpha~=~1/2$ however we may imagine an alternative dependence on time.

In the above we saw that, in an idealised version of the scheme involving perfect CNOTs, by bringing in $N-1$ additional ancilliary qubits, applying CNOTs and measuring each, that the statistics of the sum of outcomes is precisely the convolution of the statistics of an individual measurement, as given in \eqref{eqn:conv}. Here, while general distributions $\mathcal{P}_{\ket{j},t}^{(1)}$ do not convolve to distributions of the same form, for perfect CNOTs we nevertheless have  the fundamental property that the expectation and (due to independence) variance add,
${\expV{X}{\mathcal{P}_{\ket{j},t}^{(N)}} = N\expV{X}{\mathcal{P}_{\ket{j},t}^{(1)}}}$, 
${\varV{X}{\mathcal{P}_{\ket{j},t}^{(N)}} = N\varV{X}{\mathcal{P}_{\ket{j},t}^{(1)}}}$.
It therefore follows that the SNR, using the scheme, will be
\begin{equation}
    S\left(\mathcal{P}_{\ket{0},t}^{(N)},\mathcal{P}_{\ket{1},t}^{(N)}\right) = \sqrt{N}S\left(\mathcal{P}_{\ket{0},t}^{(1)},\mathcal{P}_{\ket{1},t}^{(1)}\right) \sim t^\alpha\sqrt{N}. \nonumber
\end{equation}
Thus, as long as $t^\alpha \sqrt{N}$ is constant, then the quality of the measurement is constant, and hence we still obtain a space-time trade-off. In particular, if $t_S^{(1)}$ is the duration of the measurement which achieves an SNR of $S$, then using the scheme, we can speed this up to a measurement time of $t^{(N)}_S$ given by
${t_S^{(1)}/t^{(N)}_S = N^{1/(2\alpha)}}$. This recovers \eqref{eqn:perfect_ratio} in the case $\alpha = 1/2$ but shows more generally, if $\alpha > 1/2$, we will still achieve a trade-off between space and time using this scheme, albeit not as good as \eqref{eqn:perfect_ratio}. Nevertheless, if $\alpha < 1/2$, we will achieve a ``super-improvement'' where the trade-off is better than \eqref{eqn:perfect_ratio}.

\subsection{Space-time trade-off in more realistic settings}
For our scheme to be useful in practical applications it is crucial to show that the speed-up is robust to realistic imperfection. To this end, we now describe a simple noise model to take into account the dominant sources of error. Since ancilla qubits are initialised in state $\ket{0}$, imperfections in the CNOT gate operations can be modelled by considering errors on state $\ket{10}$ being mapped to state $\ket{00}$ with error probability $p$ (see eg. \cite{gustiani2023virtualquantumdevicevqd}). Additionally, we consider imperfections in qubit readout caused by qubits in state $\ket{1}$ falling into $\ket{0}$ during the measurement, such as due to depumping by off-resonant scattering in atomic fluorescence measurements \cite{PhysRevLett.119.180504}. This error is included by applying a probabilistic decay at constant rate $\lambda$ during the measurement. See the Appendix for further details on our noise model.

The effects are summarised in Figs.~\ref{fig:snr_perfect_and_noise}c~and~\ref{fig:snr_perfect_and_noise}d. We see that we still obtain a significant space-time trade-off, allowing for significantly faster measurements, even in the presence of noise, demonstrating the robustness of this scheme. 

Fig.~\ref{fig:snr_perfect_and_noise} was generated using \eqref{eqn:snrn_noise_app}, in the Appendix, which gives a general expression for the SNR after using our scheme. It is applicable to essentially any arbitrary time-dependent noisy measurement process and accounts for arbitrary CNOT imperfections. We used \eqref{eqn:snrn_noise_app} with the noise model outlined immediately above but, by using \eqref{eqn:snrn_noise_app}, the analogue of Fig.~\ref{fig:snr_perfect_and_noise} could be reproduced much more generally for any specific architecture.

We note that our discussions have been framed in terms of the SNR, but we find essentially the same space-time trade-off if we consider the measurement infidelity: this is discussed in the Supplemental Material. In the perfect case this is shown analytically while for the imperfect case we present numerical results analogous to Fig.~\ref{fig:snr_perfect_and_noise}c and  \ref{fig:snr_perfect_and_noise}d.

There are other notable points worth highlighting. First, unlike in the perfect case, the SNR does not increase indefinitely. It reaches a peak before starting to decrease, due to decoherence. We see that the peak value is increased by schemes using more qubits. Thus not only does the scheme speed up the measurement while keeping the quality constant, it can also improve the measurement at the same time. This is shown in Fig.~\ref{fig:snr_noisy_experimental_highest_achieveable} in the Appendix in more detail. Similar to the perfect case, the rate of SNR increase improves with $N$, but it can now be a ``super-improvement''; for small enough values of $N$ and CNOT failure rate $p$, the rate of speed up is better than linear \eqref{eqn:perfect_ratio} for the relevant time scales, that is the times before decoherence causes the measurement to deteriorate, as demonstrated by Fig.~\ref{fig:snr_noisy_experimental_speed_up} in the Appendix.

\subsection{Conclusion}
In this paper we have presented a general scheme for speeding up quantum measurement while maintaining quality. This is achieved by means of a space-time trade-off, using additional ancillary qubits in place of longer measurement time. We presented the main idea using ideal Poisson measurement statistics but emphasise that the scheme is much more general than this. Moreover, by using characteristic noise models for the CNOTs and decoherence, the scheme was shown to be robust to imperfections while simultaneously able to achieve a ``super-improvement'' in speed-up, since it acts to suppresses decoherence, allowing for a higher achievable maximum quality. While we have illustrated the scheme with systems of qubits, it should be clear from the discussions above that the scheme is equally applicable to higher dimensional systems. 

Our case study was framed around state dependent fluorescence qubit measurements which are particularly appropriate to trapped-ion and cold-atom qubits. We would like to reiterate that our ideas are much more general, as demonstrated by equation \eqref{eqn:snrn_noise_app} in the Appendix. For example, our ideas also apply, when using multiple detectors, to superconducting qubits. We give a very brief discussion of this in the Supplemental Material. We also believe, just as in the earlier discussions, that we can embody our scheme for superconductors using a single detector; this is the subject of future work.

We have presented our results under the assumption that the CNOT duration is negligibly short compared to the measurement duration. The extent to which this is true is hardware specific, but it is straightforward to incorporate CNOT duration and, if it is sufficiently short, see that a trade-off will still be obtained. 
The relatively short two-qubit gate times compared to readout times in the following references demonstrate the advantage of our scheme to superconducting \cite{acharya2024quantumerrorcorrectionsurface, Bengtsson_2024} and neutral atom \cite{Bluvstein_2023, Evered_2023} quantum computers while for trapped-ion quantum computers \cite{loschnauer2024scalablehighfidelityallelectroniccontrol, Myerson_2008} the advantage is less clear cut at present. We note however that this comparison is only relevant at the time of writing. Due to the fast development of quantum technologies the precise relation between two-qubit gate and readout durations is highly likely to change in the foreseeable future. Our view is that hardware providers will need to use every tool available to make their machines as effective as possible and our scheme has the potential to become a standard part of this toolbox.

One of the most exciting applications of our scheme is to mid-circuit measurements, such as those used in error correction. As an important example, some form of parity measurement is key to many error correction schemes: here the parity of the qubits that is needed can be found by coupling those qubits to an ancilla using CNOTs, then measuring the ancilla.  This ancilla measurement, and hence the whole error detection, can be sped-up by our scheme.

Since speeding up quantum measurements is a highly relevant problem, we envisage that our scheme will have significant applications across quantum technologies.

\emph{Note added.}----During preparation of this manuscript we became aware of the independent work of Petrosyan, Norrell, Poole and Saffman \cite{PNPS24} which contains related ideas.
\begin{acknowledgments}

\section{Acknowledgements} 
We acknowledge support from the EPSRC through the following grants: EP/T005386/1,
EP/T001062/1, 
EP/Y005058/2,
EP/Y004655/1,
EP/L015730/1 and
EP/R043957/1. 
We also acknowledge support from Agence Nationale de la Recherche (ANR-22-PNCQ-0002), the Royal Society (URF NFQI) and from a CIFAR Azrieli Global Scholarship.

\end{acknowledgments}

\setcounter{equation}{0}
\renewcommand{\theequation}{A\arabic{equation}}
\subsection{Appendix A: On noise model}
We consider a noise model with two key ingredients: (i) Noise in the CNOT gates applied; (ii) decoherence of the individual systems. Initially we consider simple, albeit well-motivated models for both, but later in the Appendix we explain how the analysis can be extended to very general noise models. We first outline here these two ingredients separately in the simple model case. 

(i) For the CNOT noise, we consider a simple model inspired by neutral-atom quantum computers. In this model, whenever the system is in state $\ket{1}$ and a CNOT is applied, there is a chance of failure, which causes the system to decay to $\ket{0}$. More formally, we define a probability $p$, that immediately before a CNOT is implemented, $\ket{1}$ will be sent to $\ket{0}$. If this happens, the CNOT can not flip $\ket{0}$ back to $\ket{1}$.
The noisy CNOT channel $\widetilde{C_p}$ acts on the basis states, where the first qubit is the control qubit, as
\begin{align}\label{eqn:noisy_cnot}
    \widetilde{C_p}(\ket{00}\bra{00}) &= \ket{00}\bra{00} \nonumber\\
    \widetilde{C_p}(\ket{10}\bra{10}) &= p\ket{00}\bra{00} + (1-p)\ket{11}\bra{11}.
\end{align}

Note that since we are interested in the SNR as our figure of merit, we only need to consider the input states $\ket{\psi}=\ket 0$ and $\ket{\psi}=\ket 1$ of the system qubit. Furthermore, we assume our ancillary qubits are initialised perfectly in $\ket{0}$ so we are only interested in how the noisy CNOT gates will act on the states $\ket{00}$ and $\ket{10}$. Thus \eqref{eqn:noisy_cnot} defines the action on all states that occur in the scheme.

When the system is in the state $\ket{\psi}=\ket{0}$, and we have $N-1$ ancillary qubits, the initial state will be $\ket{0}^{\otimes N}$. In this case, the action of $\widetilde{C_p}$ is to leave the state of all $N$ systems as $\ket{0}^{\otimes N}$. When the system is in the state $\ket{\psi}=\ket{1}$, the initial state is $\ket{1}\ket{0}^{\otimes (N-1)}$. Since our scheme only depends upon the total number of photons emitted from each qubit (and does not depend upon which qubit emitted which photon), we only need to calculate the probability distribution for the total number of qubits in state $\ket{1}$ after applying the $\widetilde{C_p}$ gates. 

\begin{figure}[t!]
    \centering
    \includegraphics[width=1\linewidth]{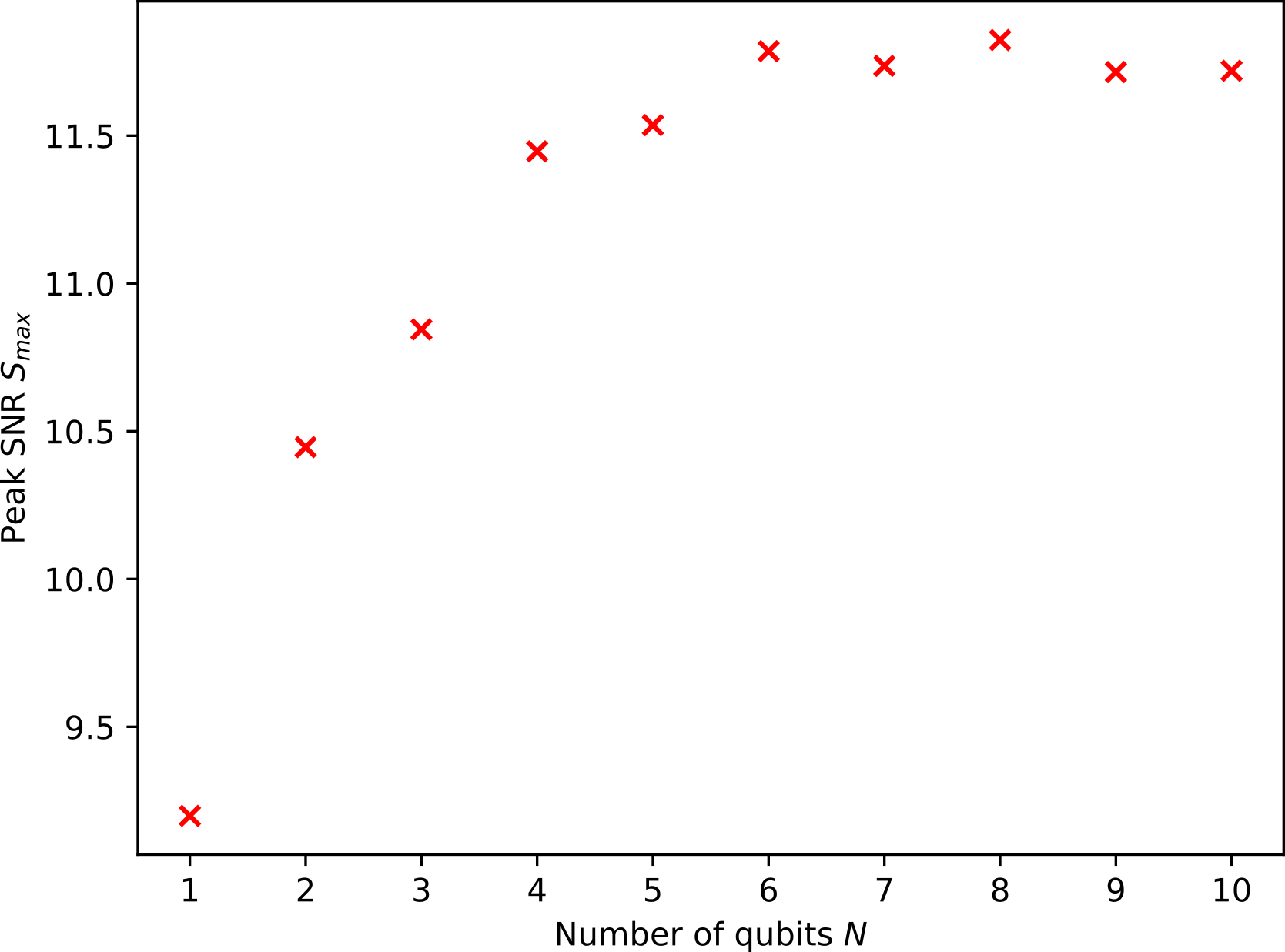}
    \caption{\textbf{Increase in peak SNR from Fig.~\ref{fig:snr_perfect_and_noise}c and \ref{fig:snr_perfect_and_noise}d.} This figure is extra detail for Figs.~\ref{fig:snr_perfect_and_noise}c~and~\ref{fig:snr_perfect_and_noise}d and uses the same parameters. We plot the peak SNR $S_{max}$ for $N=1$ to $N=10$ qubits. While more qubits results in a higher peak SNR, this trend flattens out for higher qubit numbers before eventually dropping. The slight unevenness in the plot is an odd/even effect of the cascade CNOT compilation, see the Supplemental Material.
    }
    \label{fig:snr_noisy_experimental_highest_achieveable}
\end{figure}

\begin{figure}[h!]
    \centering
    \includegraphics[width=1\linewidth]{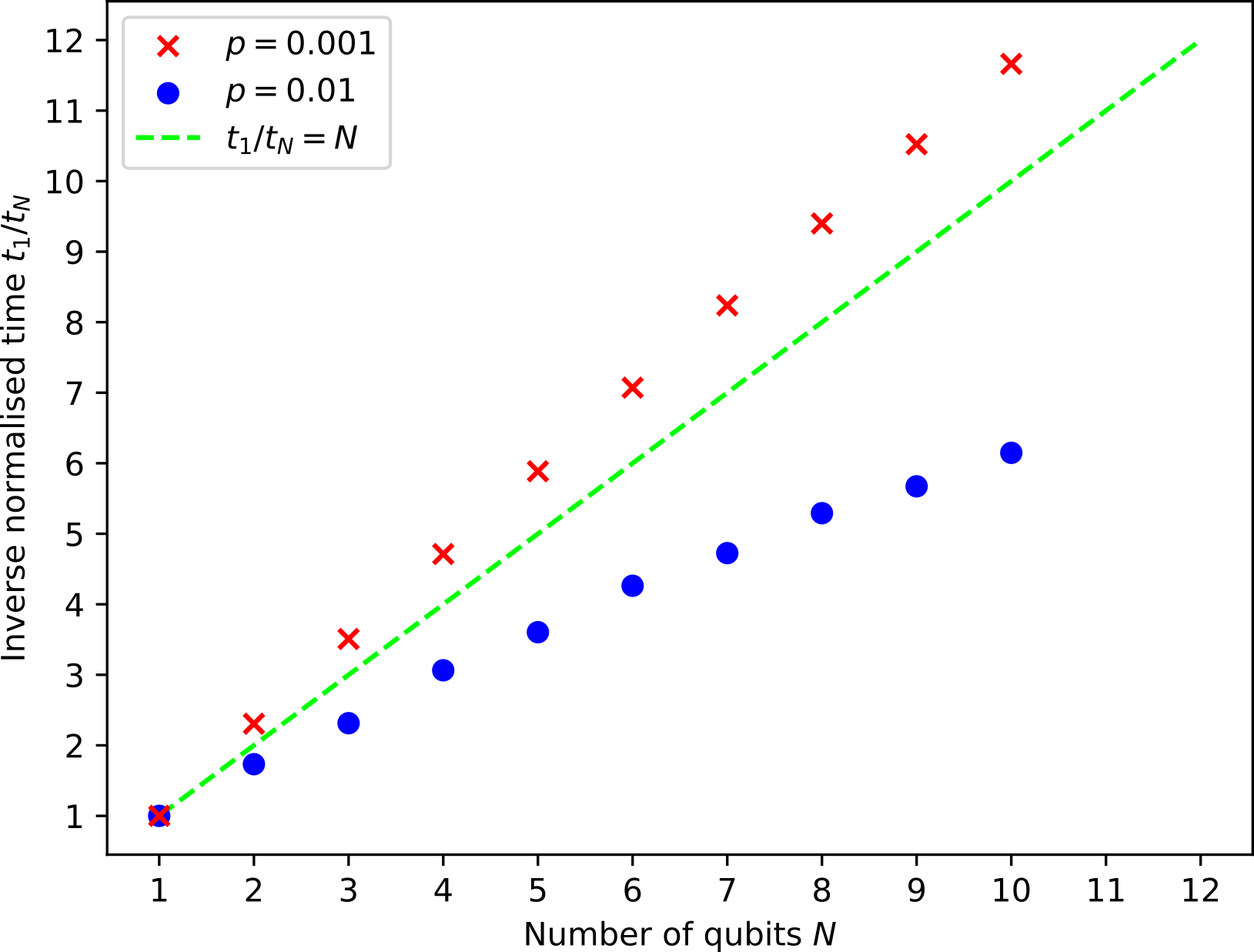}
    \caption{\textbf{Speed up from Fig.~\ref{fig:snr_perfect_and_noise}c and \ref{fig:snr_perfect_and_noise}d.} This figure is extra detail for Figs.~\ref{fig:snr_perfect_and_noise}c~and~\ref{fig:snr_perfect_and_noise}d and uses the same parameters except for $p$, for which we plot two values. We plot the ratio $t^{(1)}_S/t^{(N)}_S$ for two different values of $p$, at $S=8$, for $N=1$ to $N=10$ qubits and the linear relationship $t^{(1)}_S/t^{(N)}_S=N$ (lime dashed line) achieved by the perfect Poisson distribution \eqref{eqn:perfect_ratio}. The value $p=0.01$ (blue dots) is used in Figs.~\ref{fig:snr_perfect_and_noise}c~and~\ref{fig:snr_perfect_and_noise}d. The points are increasing, which shows we get a speed up, however they are below the dashed line so the speed-up is worse than the linear case. On the other hand, the values for $p=0.001$ (red crosses) are above the dashed line which demonstrates that, for CNOTs with a low enough failure rate, we can beat the decoherence of the qubit; it is possible to achieve a ``super-improvement''.}
    \label{fig:snr_noisy_experimental_speed_up}
\end{figure}

There are numerous ways to compile the gates and different compilations give different probability distributions for the number of qubits in state $\ket{1}$, given an input state $\ket{\psi}=\ket{1}$. These compilations are explored in more detail in the Supplemental Material. Let us denote by $\mathcal{T}^{(N)}_p(q)$ the probability distribution of the random variable $Q^{(N)}_p$ defining the number of qubits in state $\ket{1}$ after applying the noisy CNOT gates. Any qubits not in state $\ket{1}$ are in state $\ket{0}$. Combining this with the probability distribution for $N$ qubits emitting $k$ photons \eqref{eqn:conv}, we find the resultant distribution using our scheme with noisy CNOT gates to be
\begin{align}\label{eqn:noisy_cnot_protocol}
    \mathcal{P}_{\ket 0,t}^{(N)}(k) &= L_{N\mu_0t}(k), \nonumber \\ 
      \mathcal{P}_{\ket 1,t}^{(N)}(k) &= \sum^N_{q=0} \mathcal{T}^{(N)}_p(q)(L_{\mu_1t}^{*q}*L_{\mu_0t}^{*N-q})(k) \nonumber \\
      &= \sum^N_{q=0} \mathcal{T}^{(N)}_p(q)L_{(q\mu_1+(N-q)\mu_0)t}(k).
\end{align}

(ii) For the decoherence, we consider a model whereby qubits in state $\ket{1}$ probabilistically decay at a constant rate $\lambda$ to state $\ket{0}$. After this happens, the qubit will emit photons at a different rate so the resultant distribution of photons will not be the same as a perfect Poisson. We are interested in the probability distribution for the number of photons emitted $k$ in time $t$. We model this behaviour by finding the probability that a qubit in state $\ket{1}$ will emit $k'$ photons in time $t'$ before decaying to $\ket{0}$ and then emitting $k-k'$ photons in time $t-t'$, then integrating over $t'$ and summing over $k'$. We find the resultant probability distribution to be
\begin{multline}\label{eqn:decaying_poisson}
    \mathcal{W}_{\mu_1,\mu_0,\lambda,t}(k) = e^{-\lambda t}e^{-\mu_1 t}\frac{(\mu_1t)^k}{k!}\\
    +\int_0^{t}dt'\lambda e^{-\lambda t'}e^{-(\mu_1t' + \mu_0(t-t'))}\frac{(\mu_1t' + \mu_0(t-t'))^k}{k!}.
\end{multline}
The final integral can be readily evaluated numerically. 

Our complete noise and decoherence model consists of combining the noisy CNOT protocol \eqref{eqn:noisy_cnot_protocol} with the decaying Poisson distribution \eqref{eqn:decaying_poisson}. Altogether, we find that the probability distributions for the number of photons emitted, using an $N$ qubit scheme, is given by
\begin{align}\label{eqn:noise_and_decay}
    \mathcal{P}_{\ket 0,t}^{(N)}(k) &= L_{N\mu_0t}(k), \nonumber \\ 
      \mathcal{P}_{\ket 1,t}^{(N)}(k) &= \sum^N_{q=0} \mathcal{T}^{(N)}_p(q)(\mathcal{W}_{\mu_1,\mu_0,\lambda,t}^{*q}*L_{(N-q)\mu_0t})(k).
\end{align}
These are used to produce the results in Figs.~\ref{fig:snr_perfect_and_noise}c~and~\ref{fig:snr_perfect_and_noise}d where we use the cascade CNOT compilation, found in the Supplemental Material, for $\mathcal{T}^{(N)}_p(q)$.

\setcounter{equation}{0}
\renewcommand{\theequation}{B\arabic{equation}}
\subsection{Appendix B: On general space-time trade-off}
We now show how the scheme can be extended to very general noise/decoherence models with minimal assumptions. We allow for the CNOT gates to induce an arbitrary trace preserving quantum channel (i.e. we assume that the system does not leave the space spanned by the states $\ket 0$ and $\ket 1$ for each qubit) and we also assume that the measurement statistics are identical for each qubit. The SNR then depends only on (i)
$P_{\ket j ,t}(k)$, the probability distribution of the random variable $K_{\ket j,t}$ defining the measurement statistics of the basis state $\ket j$ at time $t$ for an arbitrary time-dependent noisy measurement process and (ii) $T_{\ket j}^{(N)}(q)$, the probability distribution of the random variable $Q_{\ket j}^{(N)}$ defining the final number of qubits in state $\ket j$ after applying the noisy CNOT gates to the input state $\ket j$ and $N-1$ ancillary qubits initialised in state $\ket{0}$. We note that the SNR will only depend on $T_{\ket j}^{(N)}(q)$, which is a limited statistic of the entangling CNOT process, so in order to compute the SNR we do not need full tomography of the quantum channel.

In the Supplemental Material we show that the SNR is given by
\begin{align}\label{eqn:snrn_noise_app}
    \mathcal{S}^{(N)}\left(P_{\ket{0},t}, P_{\ket{1},t}\right)=& 
    \frac{2\abs{\expV{K}{P_{\ket{1},t}}-\expV{K}{P_{\ket{0},t}}}}{\sqrt{\Delta K^2_{\mathcal{P}^{(N)}_{\ket{0},t}}}+\sqrt{\Delta K^2_{\mathcal{P}^{(N)}_{\ket{1},t}}}} \nonumber \\
    &\times \abs{\expV{Q}{\mathcal{T}_{\ket{0}}^{(N)}}+\expV{Q}{\mathcal{T}_{\ket{1}}^{(N)}}-N}
\end{align}
where  
\begin{align}
    \varV{K}{\mathcal{P}_{\ket j,t}^{(N)}}=&\expV{Q}{\mathcal{T}_{\ket{j}}^{(N)}}\varV{K}{P_{\ket{j},t}}\nonumber \\
    &+\left(N-\expV{Q}{\mathcal{T}_{\ket{j}}^{(N)}}\right)\varV{K}{P_{\ket{\overline{j}},t}}\nonumber \\
    &+\left(\expV{K}{P_{\ket{j},t}}-\expV{K}{P_{\ket{\overline{j}},t}}\right)^2\varV{Q}{\mathcal{T}_{\ket{j}}^{(N)}} \nonumber
\end{align}
and $\overline{j}=j+1 ~ {\rm mod}~2$.

Equation \eqref{eqn:snrn_noise_app} is very general. Given the time dependent measurement statistics for a single qubit $P_{\ket{j},t}(k)$ and the CNOT noise as described by the probability distributions $\mathcal{T}_{\ket{j}}^{(N)}(q)$, it is possible to calculate the SNR for an $N$ qubit scheme at any time $t$. For example, experimentally we could collect the time dependent measurement statistics of a qubit in the state $\ket{j}$ to find $P_{\ket{j},t}(k)$ and perform a - relatively limited - version of tomography on the CNOT gates to find the relevant noise distributions $T_{\ket{j}}^{(N)}(q)$. We could then use \eqref{eqn:snrn_noise_app} to generate similar plots to Fig. \ref{fig:snr_perfect_and_noise} which can be analysed to optimise over measurement duration, number of qubits and SNR.

The full derivation and further discussion of \eqref{eqn:snrn_noise_app} can be found in the Supplemental Material.

\bibliography{main}

\setcounter{equation}{0}
\renewcommand{\theequation}{S\arabic{equation}}
\renewcommand{\thefigure}{S\arabic{figure}}

\newpage\onecolumngrid
\section{Supplemental Information}

\subsection{Note on references}
Figure and equation references prefixed with S refer to the Supplemental Material. If a reference has another prefix or no prefix then it refers to the main text.

\subsection{Possible compilations of noisy CNOT gates and their respective outcome distributions}\label{app:cnot_compilation}
\begin{figure}[b!]
\centering
\includegraphics[width=0.7\linewidth]{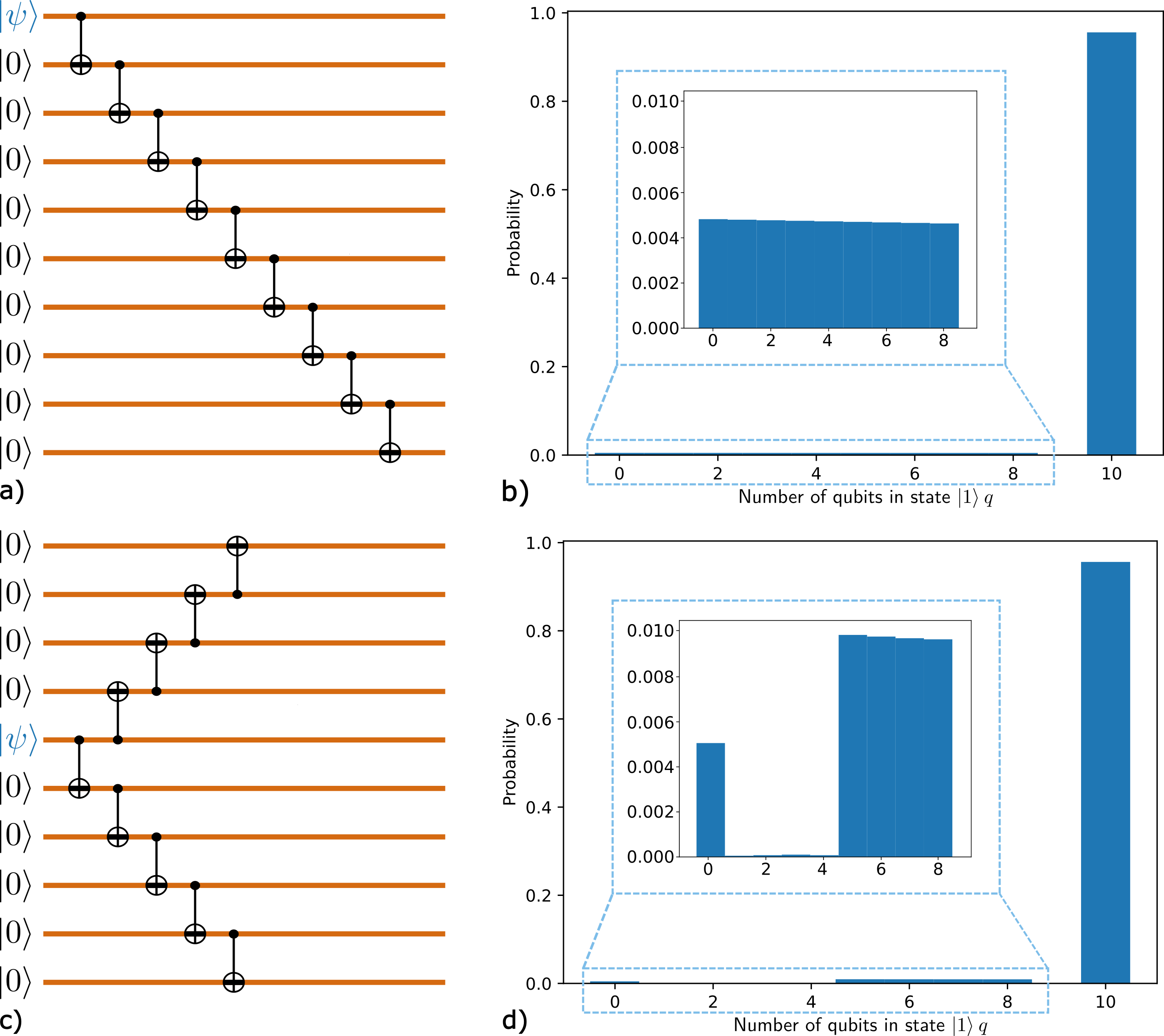}
\caption{\textbf{Flat and cascading compilations of noisy CNOT gates along with their outcome probability distributions.} All CNOTs depicted are noisy and denoted by $\widetilde{C}_p$ \eqref{eqn:noisy_cnot} where $p=0.005$. \textbf{a)} flat CNOT compilation for $N=10$ qubits. \textbf{b)} resultant probability distribution for the number of qubits in state $\ket{1}$, given the input state $\ket{\psi}=\ket{1}$, after running the flat compilation. Inset: zoomed in plot of relevant area. \textbf{c)} cascade CNOT compilation for $N=10$ qubits. The cascade compilation should in principle run much faster than the flat compilation as it is possible to run multiple CNOTs in parallel. \textbf{d)} the resultant probability distribution, along with inset, as described for \textbf{b)}, but for the cascade compilation. The probability of a complete success on all CNOTs is $~0.956$ in both cases so is much greater than any other outcome. Nevertheless, for other outcomes the cascade probability distribution is more heavily weighted towards a higher number of qubits in state $\ket{1}$ (albeit with a slightly higher chance of returning $0$ qubits in state $\ket{1}$, but this proves to be inconsequential). The probability of returning $9$ qubits in state $\ket{1}$ is $0$ for both compilations.} 
\label{fig:cnot_compilation}
\end{figure}

For a system of a single input qubit in an arbitrary state $\ket{\psi}$ and $N-1$ ancillary qubits initialised in state $\ket{0}$, we discuss some possible noisy CNOT compilations. The noisy CNOT gate $\widetilde{C_p}$ is defined in \eqref{eqn:noisy_cnot}, where $p$ is the probability of failure.
Assuming the qubits have only nearest neighbour connectivity, two compilations are: (i) flat and (ii) cascading. The flat compilation has a CNOT acting between each qubit consecutively while the cascading compilation branches into two before applying CNOTs consecutively. Fig.~\ref{fig:cnot_compilation} shows the flat and cascade compilations for 10 qubits, along with their respective probability distribution's for the number of qubits in state $\ket{1}$, denoted $q$, post implementing the CNOTs on input state $\ket{\psi}=\ket{1}$ with $p=0.005$.

The cascading compilation is superior to the flat compilation for two reasons. Firstly, since it branches into two, multiple CNOTs can be implemented in parallel, speeding up the implementation. Secondly, even though both compilations have the same chance of complete success, $(1-p)^{N-1}$, on the event of a failure, the cascading probability distribution is more heavily weighted towards a higher number of final qubits in state $\ket{1}$, since it acts to decouple errors and therefore stops them feeding forward. 

If all-to-all connectivity is possible then an obvious improvement is an exponential compilation where, at each time step, all qubits which have already been acted on by a CNOT will then act on a qubit which has yet to be acted on. This will lead to an exponential speed up in implementation time and larger decoupling of errors. Since however all-to-all connectivity is expensive, in what follows we will focus on the more realistic case of a cascaded protocol.

In what follows we define
\begin{equation}
    \theta(\mathtt{condition}) = \begin{cases}
                                1 ,& \text{if}\; \mathtt{condition}\; \text{is True}\\
                                0,& \text{if}\; \mathtt{condition}\; \text{is False}.
                            \end{cases}
\end{equation}

(i) The flat compilation probability distribution can be found by induction
\begin{equation}
    \mathcal{T}^{(N)}_{flat,p}(q) = (1-p)^q p \theta(0\leq q \leq N-2) + (1-p)^{N-1} \theta(q=N).
\end{equation}

(ii) The cascade compilation probability distribution has 3 cases, $N$ is even, $N>1$ is odd and $N=1$. If $N=2\widetilde{N}$ is even then, after the first CNOT, the cascade compilation is just two independent flat compilations of equal size. We can find this by adding the outcome of the first CNOT failing with the outcome of the first CNOT succeeding combined with the convolution of two flat compilations of size $N/2$. This results in
\begin{align}
    \mathcal{T}^{(N=2\widetilde{N})}_{cascade,p}(q) &= (1-p)(\mathcal{T}^{(\widetilde{N})}_{flat,p}*\mathcal{T}^{(\widetilde{N})}_{flat,p})(q) + p\theta(q=0), \nonumber \\
    &= (1-p)^{q+1} p^2 ((q+1)\theta(0\leq q\leq \widetilde{N} - 2) +(2\widetilde{N} - 3 - q)\theta(\widetilde{N} - 1 \leq q \leq 2(\widetilde{N} - 2))) \nonumber\\
    &\quad+ 2(1-p)^q p \theta(\widetilde{N}\leq q \leq 2\widetilde{N}-2) + (1-p)^{2\widetilde{N}-1}\theta(q=2\widetilde{N}) + p\theta(q=0).
\end{align}
If $N=2\widetilde{N}+1$ is odd and $\widetilde{N}>0$ then we add another qubit to one end of the ancillary qubits to find the result
\begin{align}\label{eqn:cascade_odd}
    \mathcal{T}^{(N=2\widetilde{N}+1)}_{cascade,p}(q) &= (1-p)(\mathcal{T}^{(\widetilde{N}+1)}_{flat,p}*\mathcal{T}^{(\widetilde{N})}_{flat,p})(q) + p\theta(q=0), \nonumber \\
    &= (1-p)^{q+1} p^2 ((q+1)\theta(0\leq q\leq \widetilde{N} - 2) + (2\widetilde{N} - 3 - q)\theta(\widetilde{N} - 1 \leq q \leq 2(\widetilde{N} - 2)) \nonumber \\
    &\quad+ \theta(\widetilde{N}-1 \leq q \leq 2\widetilde{N}-3)) + (1-p)^q p (2\theta(\widetilde{N}+1\leq q \leq 2\widetilde{N}-1) + \theta(q=\widetilde{N})) \nonumber \\
    &\quad+ (1-p)^{2\widetilde{N}}\theta(q=2\widetilde{N}+1) + p\theta(q=0).
\end{align}
If $N=1$ then $\mathcal{T}^{(1)}_{cascade,p}(q)=\mathcal{T}^{(1)}_{flat,p}(q)=\theta(q=1)$.

\subsection{General space-time trade-off}
In this section we will show how our scheme can be extended to very general noise/decoherence models with minimal assumptions. We allow for the CNOT gates to induce an arbitrary trace preserving quantum channel (i.e. we assume that the system does not leave the space spanned by the states $\ket 0$ and $\ket 1$ for each qubit) and we also assume that the measurement statistics are identical for each qubit.

We denote by $P_{\ket{j},t}(k)$ the probability distribution (or density) of the discrete (or continuous) random variable $K_{\ket j,t}$ defining the measurement statistics for the basis state $\ket{j}$, $j\in \{0,1\}$, at time $t$ for an arbitrary time-dependent noisy measurement process. We note that, since $P_{\ket{j},t}(k)$ are arbitrary, not only do they capture decohering measurement statistics, such as the decaying Poisson \eqref{eqn:decaying_poisson} analysed for realistic devices in the main text, but they can also capture any other arbitrarily complex noise which affects the measurement statistics.

We denote by $\mathcal{T}_{\ket j}^{(N)}(q)$ the probability distribution of the random variable $Q_{\ket j}^{(N)}$ defining the final number of qubits in state $\ket{j}$ after applying the noisy CNOT gates, using any compilation, to the input state $\ket{j}$ and $N-1$ ancillary qubits initialised in state $\ket{0}$. $\mathcal{T}_{\ket j}^{(N)}(q)$ will depend on all the parameters of the underlying noise.  We note that the SNR will only depend on $T_{\ket j}^{(N)}(q)$ which is a limited statistic of the entangling CNOT process, so in order to compute the SNR we do not need full tomography of the quantum channel.    With the CNOT noise analysed in the main text \eqref{eqn:noise_and_decay} there was only one parameter $p$ and we only considered noise acting on the state $\ket{1}$, but here we consider the more general case.

Assuming the state dependent measurement statistics for each qubit used in the scheme are identical, then regardless of the complexity of the noise, we ultimately only need to know the probability for the final number of qubits in the states $\ket{0}$ and $\ket{1}$. $\mathcal{T}_{\ket{j}}^{(N)}(q)$ fully captures this number. 

To derive the measurement statistics for an $N$ qubit scheme we take inspiration from the noise model discussed in the main text. Generalising the result from \eqref{eqn:noise_and_decay}, we define
\begin{equation}\label{eqn:noise_and_decay2}
      \mathcal{P}_{\ket j,t}^{(N)}(k) = \sum^N_{q=0} \mathcal{T}_{\ket j}^{(N)}(q)(P_{\ket{j},t}^{*q}*P_{\ket{\overline{j}},t}^{*(N-q)})(k)
\end{equation}
where $(f^{*N})(w)$ is the $N$-fold convolution of the distribution $f(v)$ and $\overline{j}=j+1 ~ {\rm mod}~2$.

To calculate the SNR we need to calculate the expectation and variance of the above distributions. Let~${\mathcal{L}^{(N)}_{\ket{j},q,t} = P_{\ket{j},t}^{*q}*P_{\ket{\overline{j}},t}^{*(N-q)}}$. For the state $\ket{j}$, the expectation is 
\begin{align}\label{eqn:noisy_expectation}
    \expV{K}{\mathcal{P}_{\ket j,t}^{(N)}}&=\sum_{q=0}^N\mathcal{T}_{\ket{j}}^{(N)}(q) \expV{K} {\mathcal{L}^{(N)}_{\ket{j},q,t}}, \nonumber \\
    &= \sum_{q=0}^N\mathcal{T}_{\ket{j}}^{(N)}(q)\left(q\expV{K}{P_{\ket{j},t}}+(N-q)\expV{K}{P_{\ket{\overline{j}},t}}\right), \nonumber \\
    &= \expV{Q}{\mathcal{T}_{\ket{j}}^{(N)}}\expV{K}{P_{\ket{j},t}}+\left(N-\expV{Q}{\mathcal{T}_{\ket{j}}^{(N)}}\right)\expV{K}{P_{\ket{\overline{j}},t}},
\end{align}
where we have used the additive property of the expectation under convolution. The variance is
\begin{align}
    \varV{K}{\mathcal{P}_{\ket j,t}^{(N)}}&= \sum_{q=0}^N\mathcal{T}_{\ket{j}}^{(N)}(q) \expV{K^2}{\mathcal{L}^{(N)}_{\ket{j},q,t}} -\left( \sum_{q=0}^N\mathcal{T}_{\ket{j}}^{(N)}(q) \expV{K}{\mathcal{L}^{(N)}_{\ket{j},q,t}}\right) ^2, \nonumber \\
    &= \sum_{q=0}^N\mathcal{T}_{\ket{j}}^{(N)}(q) \left(\expV{K^2} {\mathcal{L}^{(N)}_{\ket{j},q,t}} - \expV{K} {\mathcal{L}^{(N)}_{\ket{j},q,t}}^2\right) + \sum_{q=0}^N\mathcal{T}_{\ket{j}}^{(N)}(q) \expV{K} {\mathcal{L}^{(N)}_{\ket{j},q,t}}^2-\left( \sum_{q=0}^N\mathcal{T}_{\ket{j}}^{(N)}(q) \expV{K}{\mathcal{L}^{(N)}_{\ket{j},q,t}}\right) ^2, \nonumber \\
    &= \sum_{q=0}^N\mathcal{T}_{\ket{j}}^{(N)}(q) \varV{K}{\mathcal{L}^{(N)}_{\ket{j},q,t}}+\sum_{q=0}^N\mathcal{T}_{\ket{j}}^{(N)}(q) \left(q\expV{K}{P_{\ket{j},t}}+(N-q)\expV{K}{P_{\ket{\overline{j}},t}}\right)^2 \nonumber \\ 
    &\quad - \left[\expV{Q}{\mathcal{T}_{\ket{j}}^{(N)}}\expV{K}{P_{\ket{j},t}}+\left(N-\expV{Q}{\mathcal{T}_{\ket{j}}^{(N)}}\right)\expV{K}{P_{\ket{\overline{j}},t}}\right] ^2.
\end{align}
Expanding out the squared terms and using the additivity of the variance under convolution we find
\begin{equation}\label{eqn:noisy_variance}
    \varV{K}{\mathcal{P}_{\ket j,t}^{(N)}}=\expV{Q}{\mathcal{T}_{\ket{j}}^{(N)}}\varV{K}{P_{\ket{j},t}}+\left(N-\expV{Q}{\mathcal{T}_{\ket{j}}^{(N)}}\right)\varV{K}{P_{\ket{\overline{j}},t}}+\left(\expV{K}{P_{\ket{j},t}}-\expV{K}{P_{\ket{\overline{j}},t}}\right)^2\varV{Q}{\mathcal{T}_{\ket{j}}^{(N)}}.
\end{equation}

Finally, putting everything together, using \eqref{eqn:snr}, \eqref{eqn:noisy_expectation} and \eqref{eqn:noisy_variance} we find that the SNR for an $N$ qubit scheme is
\begin{equation}\label{eqn:snrn_noise}
    \mathcal{S}^{(N)}\left(P_{\ket{0},t}, P_{\ket{1},t}\right) = \frac{2\abs{\expV{K}{P_{\ket{1},t}}-\expV{K}{P_{\ket{0},t}}}}{\sqrt{\varV{K}{\mathcal{P}_{\ket 0,t}^{(N)}}}+\sqrt{\varV{K}{\mathcal{P}_{\ket 1,t}^{(N)}}}}\abs{\expV{Q}{\mathcal{T}_{\ket{0}}^{(N)}}+\expV{Q}{\mathcal{T}_{\ket{1}}^{(N)}}-N}.
\end{equation}
This equation is very general. Given the time dependent measurement statistics for a single qubit $P_{\ket{j},t}(k)$ and the CNOT noise as described by the probability distributions $\mathcal{T}_{\ket j}^{(N)}(q)$, \eqref{eqn:snrn_noise} gives the SNR for an $N$ qubit scheme at a time $t$. 

$P_{\ket{j},t}(k)$ and $\mathcal{T}_{\ket j}^{(N)}(q)$ could be found experimentally using tomographic techniques. An experimentalist could collect the time dependent measurement statistics of a qubit in the state $\ket{j}$ to find $P_{\ket{j},t}(k)$. These statistics will include any noise which affects the qubits during the measurement process. While we have assumed these statistics will be identical for all qubits, as long as they are similar \eqref{eqn:snrn_noise} will be a good approximation of how the scheme will perform. 

If complete tomography has been performed on all the CNOT gates then this will be sufficient information to find $\mathcal{T}_{\ket j}^{(N)}(q)$, however, this is more information than is necessary. To find $\mathcal{T}_{\ket j}^{(N)}(q)$, we only need to know how the CNOTs act on the states $\ket{00}$ and $\ket{10}$ and further to this, we only need to know the probability amplitudes for the total number of qubits in the state $\ket{j}$. This partial tomography may be easier to experimentally determine.

With this data and \eqref{eqn:snrn_noise}, it is possible generate similar plots to Fig. \ref{fig:snr_perfect_and_noise}, which can be used to optimise measurement duration, number of qubits and SNR. The code to generate these plots is made available via a link at the end of this document.

\subsection{Space-time trade-off using measurement infidelity}
In the main text we demonstrated how our scheme is used to provide a space-time trade-off using the \emph{signal-to-noise-ratio} (SNR) \eqref{eqn:snr} as the figure of merit. We chose the SNR as it is easy to work with mathematically and is widely used by some communities, however, it is only a function of the mean and variance of the measurement statistics and is therefore independent of the threshold value $\eta$ used to classify the measurement outcome. This means it does not fully capture the measurement quality so here we will consider an alternative figure of merit, the \emph{measurement infidelity} (MI).

Let us denote by $P_{\ket j}(x)$ the probability distribution of the random variable $X_{\ket j}$ representing the measurement statistics for the state $\ket{j}$. Recall, in the section \textbf{Case study: Poisson measurement statistics} from the main text, we assign an outcome of $0$ if $X_{\ket j}<\eta$ and $1$ if $X_{\ket j}\geq\eta$. The MI is defined as
\begin{equation}\label{eqn:infidelity}
    \mathcal{F}_{\eta}(P_{\ket{0}}, P_{\ket{1}}) = \frac{\epsilon_{\ket{0}}+\epsilon_{\ket{1}}}{2} = \frac{P_{\ket{0}}(x\geq\eta)+P_{\ket{1}}(x<\eta)}{2}.
\end{equation}
where $\epsilon_{\ket{0}}$ ($\epsilon_{\ket{1}}$) is the error, i.e. the probability of incorrectly returning $1$ when the state was $\ket{0}$ (incorrectly returning $0$ when the state was $\ket{1}$), and we assume $\expV{X}{P_{\ket{0}}}<\eta<\expV{X}{P_{\ket{1}}}$ without loss of generality. We note that, while a higher SNR represents a higher quality measurement, a lower MI represents a higher quality measurement. 

We now show how our scheme still provides a space-time trade-off when considering the MI for the Poisson distribution, $L_{\omega}(k)=\frac{\omega^k}{k!}e^{-\omega}$, discussed in \textbf{Case study: Poisson measurement statistics}. We consider an ideal $N$ qubit scheme run for time $t$. For the Poisson distributions representing the measurement statistics for the state $\ket{0}$, $L_{\mu_0Nt}(k)$, and the state $\ket{1}$, $L_{\mu_1Nt}(k)$, we define $\mu_0=\alpha\mu_1$ where we can assume $0<\alpha<1$ without loss of generality. We also define the threshold value $\eta(t)=\mu_0Nt\frac{(1/\alpha-1)}{\ln(1/\alpha)}$ calculated by solving the equation $L_{\mu_1Nt}(\eta)=L_{\mu_0Nt}(\eta)$ for $\eta$ i.e. by finding the value of $k$ that has equal probability to come from $\ket{0}$ or $\ket{1}$. By noting that ${\epsilon_{\ket{0}}=L_{\mu_0Nt}(k\geq\eta(t))=e^{-\mu_0Nt}\sum_{k=k_\eta(t)}^\infty\frac{(\mu_0Nt)^k}{k!}}$ and ${\epsilon_{\ket{1}}=L_{\mu_1Nt}(k<\eta(t))=e^{-\mu_1Nt}\sum_{k=0}^{k_\eta(t)-1}\frac{(\mu_1Nt)^k}{k!}}$, where $k_\eta(t)=\ceil{\eta(t)}$, we can bound the error probabilities by
\begin{align}
    e^{-\mu_0Nt}\frac{(\mu_0Nt)^{k_\eta(t)}}{k_\eta(t)!} &\leq \epsilon_{\ket{0}} \leq e^{-\mu_0Nt}\frac{(\mu_0Nt)^{k_\eta(t)}}{k_\eta(t)!}\frac{k_\eta(t)}{k_\eta(t)-\mu_0Nt}, \nonumber \\
    e^{-\mu_1Nt}\frac{(\mu_1Nt)^{k_\eta(t)-1}}{(k_\eta(t)-1)!} &\leq \epsilon_{\ket{1}} \leq e^{-\mu_1Nt}\frac{(\mu_1Nt)^{k_\eta(t)-1}}{(k_\eta(t)-1)!}\frac{\mu_1Nt}{\mu_1Nt-(k_\eta(t)-1)},
\end{align}
where the lower bound in each case comes from considering the leading order term in the summation and the upper bound comes by bounding the summation by a geometric series. The above bounds show that the leading order behaviour is given by
\begin{align}\label{eqn:error_approx}
    L_{\mu_0Nt}(k\geq\eta(t))&\approx e^{-\mu_0Nt}\frac{(\mu_0Nt)^{k_\eta(t)}}{k_\eta(t)!}\approx e^{-\mu_0Nt\gamma_0}, \nonumber \\
    L_{\mu_1Nt}(k<\eta(t))&\approx e^{-\mu_1Nt}\frac{(\mu_1Nt)^{k_\eta(t)}}{(k_\eta(t))!} \approx e^{-\mu_1Nt\gamma_1},
\end{align}
where $\gamma_0=\beta \ln(\beta)+1-\beta$, $\gamma_1=\alpha\beta \ln(\alpha\beta)+1-\alpha\beta$ and $\beta=\frac{(1/\alpha-1)}{\ln(1/\alpha)}$. The $\gamma_j$ factors can be shown to be greater than $0$ and are calculated using the Stirling approximation and $k_\eta(t)\approx\eta(t)$. Thus both contributions to the measurement  infidelity \eqref{eqn:infidelity} decrease exponentially with $Nt$ and we see the exact space-time trade-off. 

\begin{figure}[t!]
\centering
\includegraphics[width=\linewidth]{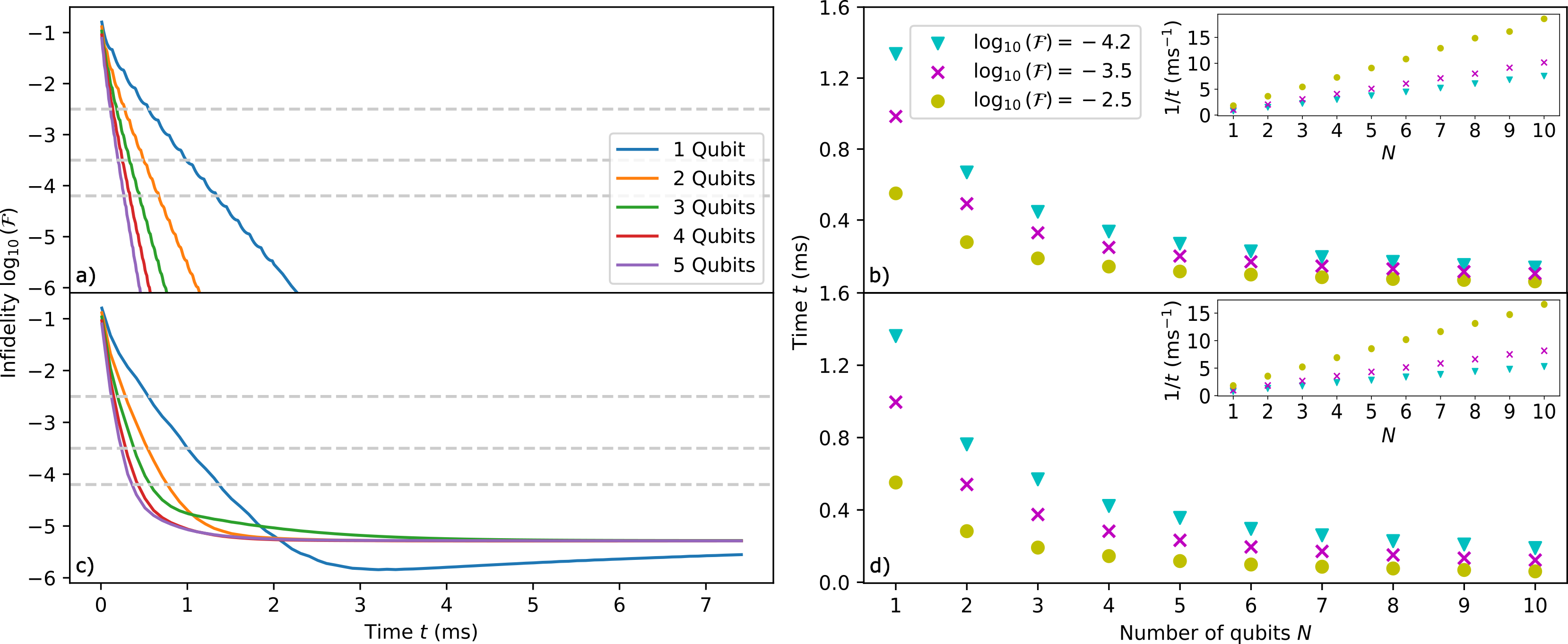}
\caption{{\bf Space-time trade-off for idealised and realistic devices using MI.} This figure is in direct comparison to Fig.~\ref{fig:snr_perfect_and_noise} which uses the SNR as the figure of merit. The parameters used are also those of Fig.~\ref{fig:snr_perfect_and_noise}. \textbf{a)} MI against time from $N=1$ to $N=5$ qubits. The MI always decreases with time and qubit number and continues the trend beyond what is shown on the graph. The small oscillations, especially apparent for the single qubit line, are present because the error rates require us to round our threshold value to the closest integer. \textbf{b)} time of measurement $t$ against number of qubits from $N=1$ to $N=10$ qubits. Inset: same points but for the inverse time $1/t$. \textbf{c)} and \textbf{d)} show similar results but for realistic devices. \textbf{c)} MI against time for different numbers of qubits $N$. The MI decreases with time down to a minimum value then increases due to decoherence. The rate at which the MI improves increases with qubit number, however, unlike with the SNR, we do not see an improvement in best MI achieved with qubit number. This is not a feature of the scheme; it is a feature of the parameters and is explored in more depth in Figs.~\ref{fig:fidelity_001p}~and~\ref{fig:fidelity_measurement_stats}. \textbf{d)} time of measurement $t$, when the defined MI is first achieved, against number of qubits $N$. Inset: same points plotted for inverse time $1/t$.} 
\label{fig:fidelity_combine}
\end{figure}

The trade-off is displayed in Fig.~\ref{fig:fidelity_combine} which is equivalent to Fig.~\ref{fig:snr_perfect_and_noise} but using the MI instead of the SNR. Figs.~\ref{fig:fidelity_combine}a~and~\ref{fig:fidelity_combine}b show the expected exponential relationship and space-time trade-off.

The above calculations were for the ideal case, without noise or decoherence. Figs.~\ref{fig:fidelity_combine}c~and~\ref{fig:fidelity_combine}d show the MI, now including noise on the gates and decoherence. We still see the space-time trade-off in this case, however, there are a few differences between the MI and the SNR. 

\begin{figure}[t!]
\centering
\includegraphics[width=0.55\linewidth]{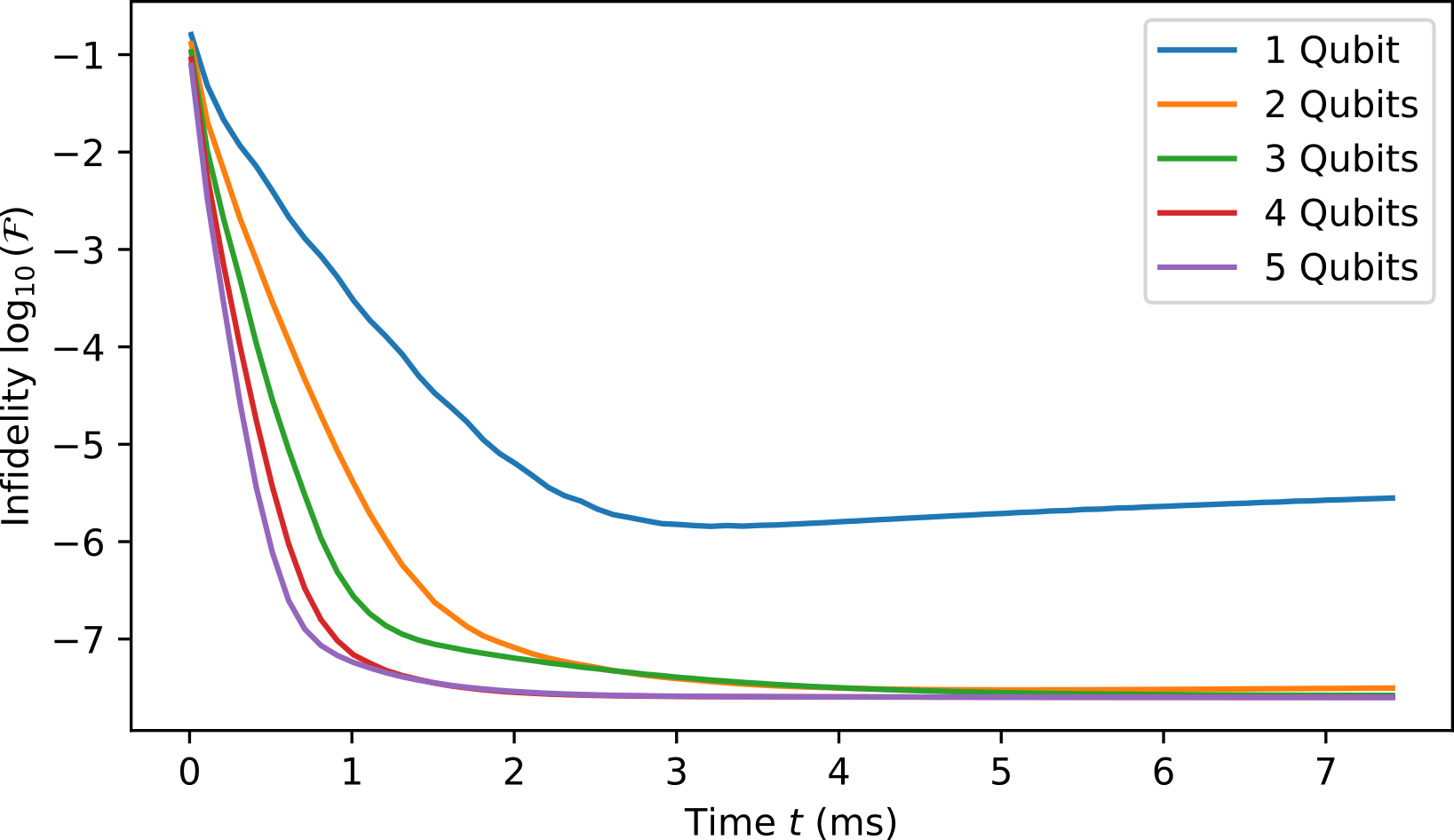}
\caption{\textbf{Space-time trade-off for realistic device using MI and low CNOT failure rate.} MI \eqref{eqn:infidelity} against time for a realistic device. The parameters used are the same as Figs.~\ref{fig:snr_perfect_and_noise}c~and~\ref{fig:snr_perfect_and_noise}d except $p=0.001$ ($10\times$ smaller). This demonstrates that, if $p$ is small enough, it is possible to achieve an improvement in the optimal MI by increasing the number of qubits. It is worth noting that, in both Figs.~\ref{fig:fidelity_combine}c~and~\ref{fig:fidelity_001p}, the optimal MI becomes constant with qubit number for $N\geq 2$. This can be explained by studying the measurement statistics and is discussed in the text.} 
\label{fig:fidelity_001p}
\end{figure}

Firstly, unlike with the SNR, where we saw an improvement in the peak value for increasing qubit numbers, we no longer see this improvement for optimal MI. However, the fact the lowest infidelity is found for 1 qubit is not a feature of the scheme, but a feature of the parameters chosen. Fig.~\ref{fig:fidelity_001p} shows the MI over time using a lower CNOT failure rate parameter, $p=0.001$ instead of $p=0.01$. For the lower CNOT failure rate we recover the improvement in optimal MI. 

\begin{figure}[h!]
\centering
\includegraphics[width=0.95\linewidth]{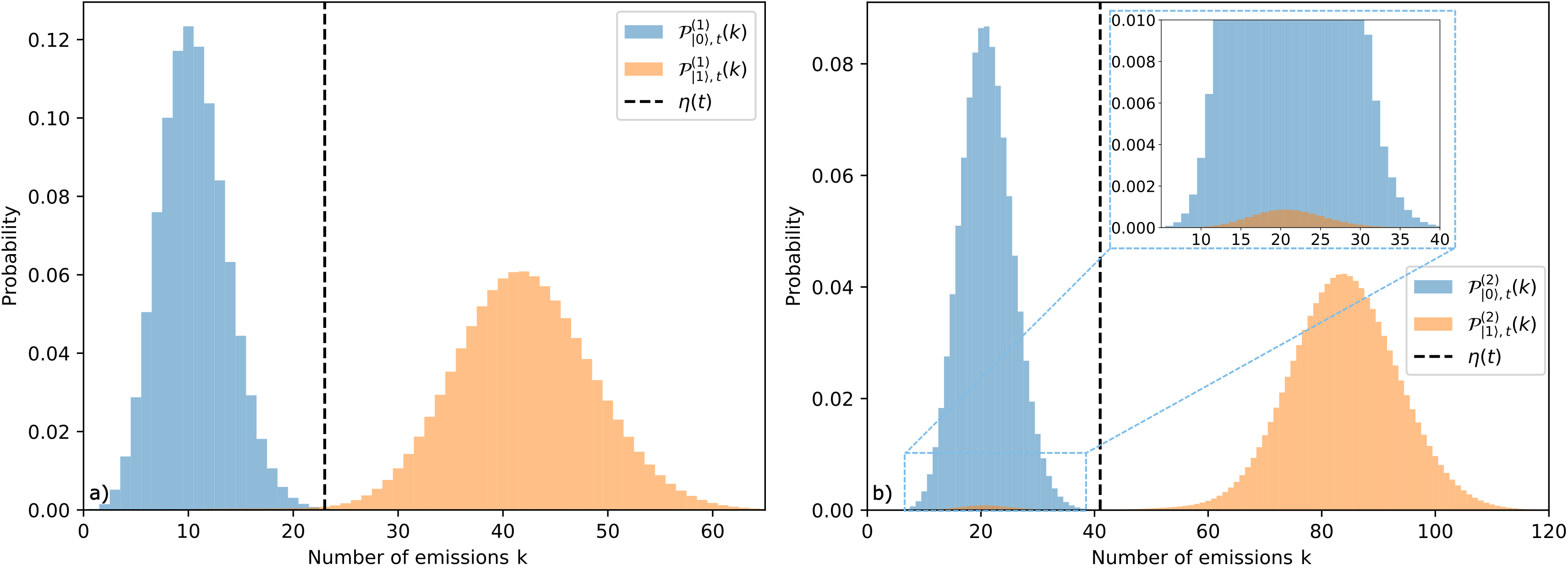}
\caption{\textbf{Measurement statistics for a realistic device.} Measurement statistics $P_{\ket{0},t}^{(N)}(k)$ and $P_{\ket{1},t}^{(N)}(k)$, at a snapshot in time $t=3\text{ms}$. The parameters used are the same as Figs.~\ref{fig:snr_perfect_and_noise}c~and~\ref{fig:snr_perfect_and_noise}d. In the context of fluorescence measurement, studied in \textbf{Case study: Poisson measurement statistics}, the horizontal axis shows the number of photon emissions $k$ while the vertical axis shows probability of measuring $k$. The dashed black line is the optimal threshold value $\eta(t)$ which minimises the MI \eqref{eqn:infidelity}. \textbf{a)} measurement statistics for $N=1$ qubit. \textbf{b)} measurement statistics for $N=2$ qubits. Inset: zoomed in plot of relevant area.}
\label{fig:fidelity_measurement_stats}
\end{figure}

Secondly, in both the $p=0.01$ and $p=0.001$ cases we note that the optimal MI appears to be approximately independent of the qubit number, when $N \geq 2$. This effect can be explained by studying the measurement statistics. Fig.~\ref{fig:fidelity_measurement_stats} shows the measurement statistics for $N=1$ and $N=2$ when $p=0.01$. While the average distance between the distributions appears to be greater for the $N=2$ case, the distribution $P_{\ket{1},t}^{(2)}(k)$ contains a small peak which lies within the distribution $P_{\ket{0},t}^{(2)}(k)$. This peak is present when we are measuring a qubit in state $\ket{1}$ using an $N\geq 2$ qubit scheme and is due to the probability of the first CNOT failing in the cascade CNOT compilation. This failure results in all further CNOTs failing and produces the state $\ket{0}^{\otimes N}$, hence the small peak lying within the distribution $P_{\ket{0},t}^{(2)}(k)$. The failure probability of the cascade CNOT compilation is illustrated in Fig.~\ref{fig:cnot_compilation}d. The small peak is approximately the same size regardless of the number of qubits and will always be on the incorrect side of the threshold value. This causes the optimal MI to be approximately constant for all qubit numbers, $N\geq 2$. On the other hand, the SNR, which is only a function of the mean and variance, is insensitive to this small peak so is able to increase with qubit number.

\subsection{Application of our scheme to superconducting qubits}
Looking beyond trapped-ions and neutral-atoms, we can also see how our protocol will work for superconducting qubits. These are most commonly measured using dispersive resonator readout \cite{Krantz_2019, Walter_2017, swiadek2023, PhysRevA.79.013819, PhysRevA.69.062320, PhysRevA.77.012112}. A qubit with frequency $\omega_a$ and a resonator with frequency $\omega_r$ are coupled with a coupling strength $g$ where the dispersive approximation $\Delta=\omega_a-\omega_r$, $|\Delta|\ll g$ is used. This coupling causes the resonator's frequency to shift by $(-1)^j\frac{g^2}{\Delta}$ depending on the qubit state $\ket{j}$. To measure the qubit, microwaves at a frequency $\omega_R$ are transmitted towards the resonator for a time $t$, which are then reflected back. The reflection causes a phase shift dependent on the frequency shift of the resonator and therefore dependent on the state of the qubit. The phase shift is measured using an IQ mixer with the result quantified as an angle on the IQ-plane. This can be achieved with either heterodyne or homodyne measurement techniques. 

For a single shot measurement on a qubit in state $\ket{j}$, a measurement device will record a measurement signal $\theta$. Let us denote by $\mathcal{P}_{\ket j,t}^{(N)}(\theta)$ the probability density of the random variable $\Theta_{\ket j,t}^{(N)}$, defining the measurement signal, when we perform an $N$ qubit scheme for time $t$ and the qubit state is $\ket j$. In the idealised case, the measurement signal is Gaussian distributed \cite{Krantz_2019, Walter_2017, swiadek2023, PhysRevA.79.013819, PhysRevA.69.062320, PhysRevA.77.012112}, in particular  \cite{PhysRevA.77.012112},
\begin{equation}\label{eqn:gaussian}
    \mathcal{P}_{\ket j,t}^{(1)}(\theta) = G_{z_j(t), \sqrt{|z_j(t)|}}(\theta),
\end{equation}
where $z_j(t)\propto (-1)^j \frac{g^2}{\Delta} t$, $G_{m, \sigma}(\theta)=\frac{1}{\sigma\sqrt{2\pi}}e^{-\frac{1}{2}(\frac{\theta-m}{\sigma})^2}$ with mean $\expV{\Theta}{G_{m, \sigma}}=m$ and variance $\varV{\Theta}{G_{m, \sigma}}=\sigma^2$. Like the Poisson case, the mean and variance are time dependent and the outcome distributions have an overlap which lowers the quality of the measurement. 

Bearing in mind we are now dealing with a continuous distribution, all of the results for Poisson statistics follow through for Gaussian statistics. In particular, the convolution of Gaussian densities is again a Gaussian density (with the sum of expectation values, and variance). We thus obtain the same results as in the section \textbf{Exact space-time trade-off in the idealised Poisson case}. We find that using an $N$ qubit scheme, we speed up the measurement of the target qubit by a factor of $N$, while maintaining the quality (as quantified by the SNR). This exact space-time trade-off does not require that the mean and variance of the Gaussian are equal (as in \eqref{eqn:gaussian}); only that they are both linear in time.

\subsection{Code}
All the code used to generate the data and plots in this paper can be found at the GitHub link: 

\url{https://github.com/custal/speeding-up-quantum-measurement}

\end{document}